\documentclass[11pt]{article}

\usepackage{amsmath,amssymb,amsfonts,amsthm,dsfont,dcolumn}
\usepackage[margin=-5pt]{caption}
\usepackage{graphicx,wrapfig}
\usepackage{tensor}
\usepackage[hypertex]{hyperref}

\DeclareGraphicsExtensions{.pdf,.png,.jpg,.jpeg}

\newcolumntype{d}{D{.}{.}{0}}

\newlength{\textlength}
\newlength{\overlinelength}

\usepackage{color}

\newcounter{subequation}[equation]

\newcommand{\be}{\begin{equation}}
\newcommand{\ee}{\end{equation}}
\newcommand{\eel}[1]{\label{#1}\end{equation}}
\newcommand{\bea}{\begin{eqnarray}}
\newcommand{\eea}{\end{eqnarray}}
\newcommand{\eeal}[1]{\label{#1}\end{eqnarray}}

\makeatletter

\def\thesubequation{\theequation\@alph\c@subequation}
\def\@subeqnnum{{\rm (\thesubequation)}}
\def\slabel#1{\@bsphack\if@filesw {\let\thepage\relax
   \xdef\@gtempa{\write\@auxout{\string
      \newlabel{#1}{{\thesubequation}{\thepage}}}}}\@gtempa
   \if@nobreak \ifvmode\nobreak\fi\fi\fi\@esphack}
\def\subeqnarray{\stepcounter{equation}
\let\@currentlabel=\theequation\global\c@subequation\@ne
\global\@eqnswtrue \global\@eqcnt\z@\tabskip\@centering\let\\=\@subeqncr

$$\halign to \displaywidth\bgroup\@eqnsel\hskip\@centering
  $\displaystyle\tabskip\z@{##}$&\global\@eqcnt\@ne
  \hskip 2\arraycolsep \hfil${##}$\hfil
  &\global\@eqcnt\tw@ \hskip 2\arraycolsep
  $\displaystyle\tabskip\z@{##}$\hfil
   \tabskip\@centering&\llap{##}\tabskip\z@\cr}
\def\endsubeqnarray{\@@subeqncr\egroup
                     $$\global\@ignoretrue}
\def\@subeqncr{{\ifnum0=`}\fi\@ifstar{\global\@eqpen\@M
    \@ysubeqncr}{\global\@eqpen\interdisplaylinepenalty \@ysubeqncr}}
\def\@ysubeqncr{\@ifnextchar [{\@xsubeqncr}{\@xsubeqncr[\z@]}}
\def\@xsubeqncr[#1]{\ifnum0=`{\fi}\@@subeqncr
   \noalign{\penalty\@eqpen\vskip\jot\vskip #1\relax}}
\def\@@subeqncr{\let\@tempa\relax
    \ifcase\@eqcnt \def\@tempa{& & &}\or \def\@tempa{& &}
      \else \def\@tempa{&}\fi
     \@tempa \if@eqnsw\@subeqnnum\refstepcounter{subequation}\fi
     \global\@eqnswtrue\global\@eqcnt\z@\cr}
\let\@ssubeqncr=\@subeqncr
\@namedef{subeqnarray*}{\def\@subeqncr{\nonumber\@ssubeqncr}\subeqnarray}

\@namedef{endsubeqnarray*}{\global\advance\c@equation\m@ne
                           \nonumber\endsubeqnarray}

\makeatletter \@addtoreset{equation}{section} \makeatother
\renewcommand{\theequation}{\thesection.\arabic{equation}}

\catcode`\@=11

\newcount\hour
\newcount\minute
\newtoks\amorpm \hour=\time\divide\hour by 60\minute
=\time{\multiply\hour by 60 \global\advance\minute by-\hour}
\edef\standardtime{{\ifnum\hour<12 \global\amorpm={am}
        \else\global\amorpm={pm}\advance\hour by-12 \fi
        \ifnum\hour=0 \hour=12 \fi
        \number\hour:\ifnum\minute<10
        0\fi\number\minute\the\amorpm}}
\edef\militarytime{\number\hour:\ifnum\minute<10 0\fi\number\minute}

\def\draftlabel#1{{\@bsphack\if@filesw {\let\thepage\relax
   \xdef\@gtempa{\write\@auxout{\string
      \newlabel{#1}{{\@currentlabel}{\thepage}}}}}\@gtempa
   \if@nobreak \ifvmode\nobreak\fi\fi\fi\@esphack}
        \gdef\@eqnlabel{#1}}
\def\@eqnlabel{}
\def\@vacuum{}
\def\marginnote#1{}
\def\draftmarginnote#1{\marginpar{\raggedright\scriptsize\tt#1}}
\def\draft{
        \pagestyle{plain}
        \overfullrule=2pt
        \oddsidemargin -.5truein
        \def\@oddhead{\sl \phantom{\today\quad\militarytime} \hfil
        \smash{\Large\sl DRAFT} \hfil \today\quad\militarytime}
        \let\@evenhead\@oddhead
        \let\label=\draftlabel
        \let\marginnote=\draftmarginnote
        \def\ps@empty{\let\@mkboth\@gobbletwo
        \def\@oddfoot{\hfil \smash{\Large\sl DRAFT} \hfil}
        \let\@evenfoot\@oddhead}

\def\@eqnnum{(\theequation)\rlap{\kern\marginparsep\tt\@eqnlabel}
        \global\let\@eqnlabel\@vacuum}  }

\renewcommand{\theequation}{\thesection.\arabic{equation}}

\def\appendix#1{
  \addtocounter{section}{-3}
  \setcounter{equation}{0}
  \renewcommand{\thesection}{\Alph{section}}
  \section*{Appendix \thesection\protect\indent \parbox[t]{11.15cm}
  {#1} }
  \addcontentsline{toc}{section}{Appendix \thesection\ \ \ #1}
  }
\def\be{\begin{equation}}
\def\ee{\end{equation}}

\date{}
\begin{document}

\begin{titlepage}

\begin{center}

{\Large \bf Turbulence and Chaos in Anti-de-Sitter Gravity}

\vskip .7 cm

\vskip 1 cm

{\large   H. P. de Oliveira$^1$, Leopoldo A. Pando Zayas$^2$, and C\'esar A. Terrero-Escalante$^3$}

\end{center}

\vskip .4cm \centerline{\it ${}^1$ Universidade do Estado do Rio de Janeiro}
\centerline{\it Instituto de F\'{\i}sica  --  Departamento de F\'{\i}sica Te\'orica}
\centerline{\it CEP 20550-013 Rio de Janeiro, RJ, Brazil.}

\vskip .4cm \centerline{\it ${}^1$ Michigan Center for Theoretical
Physics}
\centerline{ \it Randall Laboratory of Physics, The University of
Michigan}
\centerline{\it Ann Arbor, MI 48109-1120}

\vskip .4cm \centerline{\it ${}^2$  Facultad de Ciencias}
\centerline{ \it Universidad de Colima}
\centerline{\it Bernal D\'{\i}az del Castillo 340, Col. Villas San Sebasti\'an,
Colima}
\centerline{\it Colima 28045, M\'exico}

\vskip .4cm
\centerline{ \it }
\centerline{\it  }

\vskip 1 cm

\vskip 1.5 cm

\begin{abstract}
Due to the AdS/CFT correspondence the question of instability of Anti-de-Sitter spacetimes sits in the intersection of mathematical and numerical relativity, string theory, field theory and condensed matter physics. In this essay we revisit that important question emphasizing the power of spectral methods and highlighting the effectiveness of standard techniques for studying nonlinear dynamical systems. In particular we display explicitly how the problem can be modeled as a system on nonlinearly coupled harmonic oscillators. We highlight some of the many open questions that stem from this result and point out that a full understanding will necessarily required the interdisciplinary cooperation of various communities.
\end{abstract}

\vspace{2cm}

\begin{center}
{\bf Essay awarded honorable mention in the Gravity Research Foundation essay competition 2012.}
\end{center}

\end{titlepage}

The Anti-de-Sitter/Conformal Field Theory correspondence (AdS/CFT) has been responsible for the growing interest in spacetimes with negative cosmological constant $\Lambda < 0$, that is, in asymptotically locally AdS spacetimes.  The main content of the AdS/CFT correspondence is the identification of asymptotic data of fields in AdS with sources and vacuum expectation values of operators in the dual field theory
\cite{Maldacena:1997re}.

The AdS/CFT correspondence was readily expanded to include field theories at finite temperature. In this context a field theory in equilibrium at finite temperature is dual to a black hole in asymptotically AdS spacetime. One very important development has been the
establishment of the correspondence at the level of small fluctuations.  More precisely, fluctuations of systems at equilibrium or weakly driven near equilibrium satisfy a universal relation known as the fluctuation-dissipation theorem which connects spontaneous fluctuations to the linear response. The search for similar relations for systems far from equilibrium has been an important problem for many years with applications in ion collisions at Relativistic Heavy Ion Collider (RHIC) and Large Hadron Collider (LHC) and also in condensed matter physics. The AdS/CFT
correspondence has successfully incorporated those near-equilibrium relations and is now posed to take the next step and tackle far-from-equilibrium phenomena. To study far-from-equilibrium phenomena one needs to study the evolution of Einstein's equation with appropriate boundary conditions. Therefore the study of gravitational collapse has emerged as one of the most interesting problems in the interface of the interests various communities including, among others, the numerical relativity community, string theory, high energy particle physics and condensed matter.

An important property of asymptotically $AdS$ spacetimes,which is different from our intuition in asymptotically flat spacetimes, is the presence of a timelike boundary at spatial and null infinity where suitable boundary conditions need to be prescribed.

In this context the study of the dynamics of a self interacting massless scalar field in AdS spacetime has recently been initiated (see for example \cite{bizon,leo-garf}).  Even at this very incipient state very interesting results that remind us of the richness of the nonlinearities of Einstein's field equations have already been obtained. Before entering into the qualitative description of these features, let us consider the model studied by Bizon and Rostworowski (BR) \cite{bizon} in which the spacetime is described by,

\be
ds^2=\frac{\mathit{l}^2}{\cos^2x}\bigg[-A e^{-2\delta} dt^2 + A^{-1}dx^2 +\sin^2x \left(d\theta^2 +\sin^2\theta d\phi^2 \right)\bigg],
\ee

\noindent where $\mathit{l}^2=-3/\Lambda$, and the metric functions $A,\delta$ depend on $(t,x)$. Here $t$ is the time coordinate and $0 \leq x \leq \pi/2$ covers the spatial domain with the boundaries at the origin ($x=0$) and the spatial infinity ($x=\pi/2$). To write down the field equations we define the auxiliary variables $\Pi$ and $\Phi$ as,

\be
\Pi= A^{-1}e^{\delta} \dot{\varphi}, \qquad \Phi =\varphi'.
\ee

\noindent where $\varphi(t,x)$ is the scalar field, prime and dot means derivatives with respect to $x$ and $t$, respectively. Now, the Einstein-Klein-Gordon system reads,

\begin{eqnarray}
\dot{\Phi}&=& (A \mathrm{e}^{-\delta} \Pi)^\prime, \qquad \dot{\Pi}=\frac{1}{\tan^2 x}\left(A e^{-\delta} \Phi \tan^2 x \right)', \\
\delta'&=&-\sin x\cos x (\Phi^2+\Pi^2),\\
A'&=&-A (\Phi^2 +\Pi^2) \sin x\cos x +\frac{(1-A)(1+2\sin^2x)}{\sin x\cos x},
\end{eqnarray}

\noindent which show the evolution and constraint constraint. The most simple exact solution of this system is the pure AdS spacetime characterized by $\Pi=\Phi=0$, $\delta=0$ and $A=1$. As a matter of fact this solution can be viewed as analogous to the Minkowski spacetime when $\Lambda=0$. 


According to BR the scalar field evolves by performing a sequence of reflections between the boundaries and eventually is driven to instability to finally collapse forming a black hole. They have shown that this feature is valid for any arbitrarily small initial amplitude of the scalar field implying instability of AdS spacetime, contrary to the stability of the Minkowski spacetime with respect to arbitrarily small perturbations which can essentially dissipate to spatial infinity. The main reason behind such surprising result is the fact that the timelike nature of the boundaries of AdS spacetime act as a confining box. The continued reflections of the scalar field eventually excite modes with higher and higher energies that produces an inevitable road toward the instability suggesting typical signature of a turbulent system. Further supporting evidence to the claim of BR was provided in Dias,Horowitz and Santos  \cite{dias} who considered the purely gravitational problem of nonlinear stability of AdS by studying non-spherically symmetric configurations. This work was performed perturbatively and therefore could not explore the full nonlinear regime such as horizon formation. Nevertheless it gathered evidence to the claim that energy is transferred to ever higher and higher frequency modes.

In this essay we demonstrate the efficacy of spectral methods \cite{boyd} in gaining understanding of the underlying physics. For this end we have solved numerically the field equations using spectral methods for which a brief description follows. The start up and also the central idea of any spectral method is to approximate the relevant fields - $\Pi,\Phi,A,\delta$ - as appropriate series with respect to certain basis functions. For instance, taking the field $\Pi$, we have,

\begin{equation}
\Pi_a(t,y) = \sum_{k=0}^N\,a_k(t) \psi_k(y),
\end{equation}

\noindent where $a_k(t)$ are the unknown modes, the variable $y$ is a rescaling of $x$, $y=4x/\pi-1$, so that the origin $x=0$ and the spatial infinity $x=\pi/2$ are located at $y=-1,+1$, respectively. The basis functions $\psi_k(y)$ are constructed such that the conditions for smoothness at the origin, at spatial infinity and also the finiteness of the total mass-energy content of the spacetime are satisfied. This means that,

\bea
\Pi(t,y) &=& \Pi_0(t) + \mathcal{O}((y+1)^2),\qquad \mathrm{near\, y=-1} \\
\Pi(t,y) &=&  \mathcal{O}((y-1)^{-2}),\qquad \mathrm{near\, y=1}.
\eea

\noindent One of the main consequences of the spectral method is to transform a partial differential equation into a finite set of ordinary differential equations. In the present case these equations are for the modes modes $a_k(t)$.

Following BR we consider initial data of the form,
\be
\Phi(0,y)=0, \qquad \Pi(0,y)=\epsilon_0 \exp\left[-\frac{\tan^2(\pi/4(1+y))}{\sigma^2}\right]
\ee

\noindent where we have fixed $\sigma=1/4$ and $\epsilon_0$ is the initial amplitude of the scalar field distribution. As a very useful illustration of the reflections of the scalar field, we show in Fig. 1 a sequence of plots of $\Pi(t,y)$ evaluated in distinct intervals of time. The first plot correspond to the initial distribution in which we have selected $\epsilon_0=0.5$ and the other plots are evaluated in several subsequent instants. The arrow indicates the sense of propagation that is toward the boundary $y=1$ and after a reflection returns to the boundary $y=-1$. This overall periodic motion remains for a long time before the instabilities triggered by the gravitational sector begin to dominate the dynamics. The greater the  initial amplitude the less time the system undergoes this periodic motion.


\begin{figure}[htb]
\rotatebox{0}{\includegraphics*[scale=0.2]{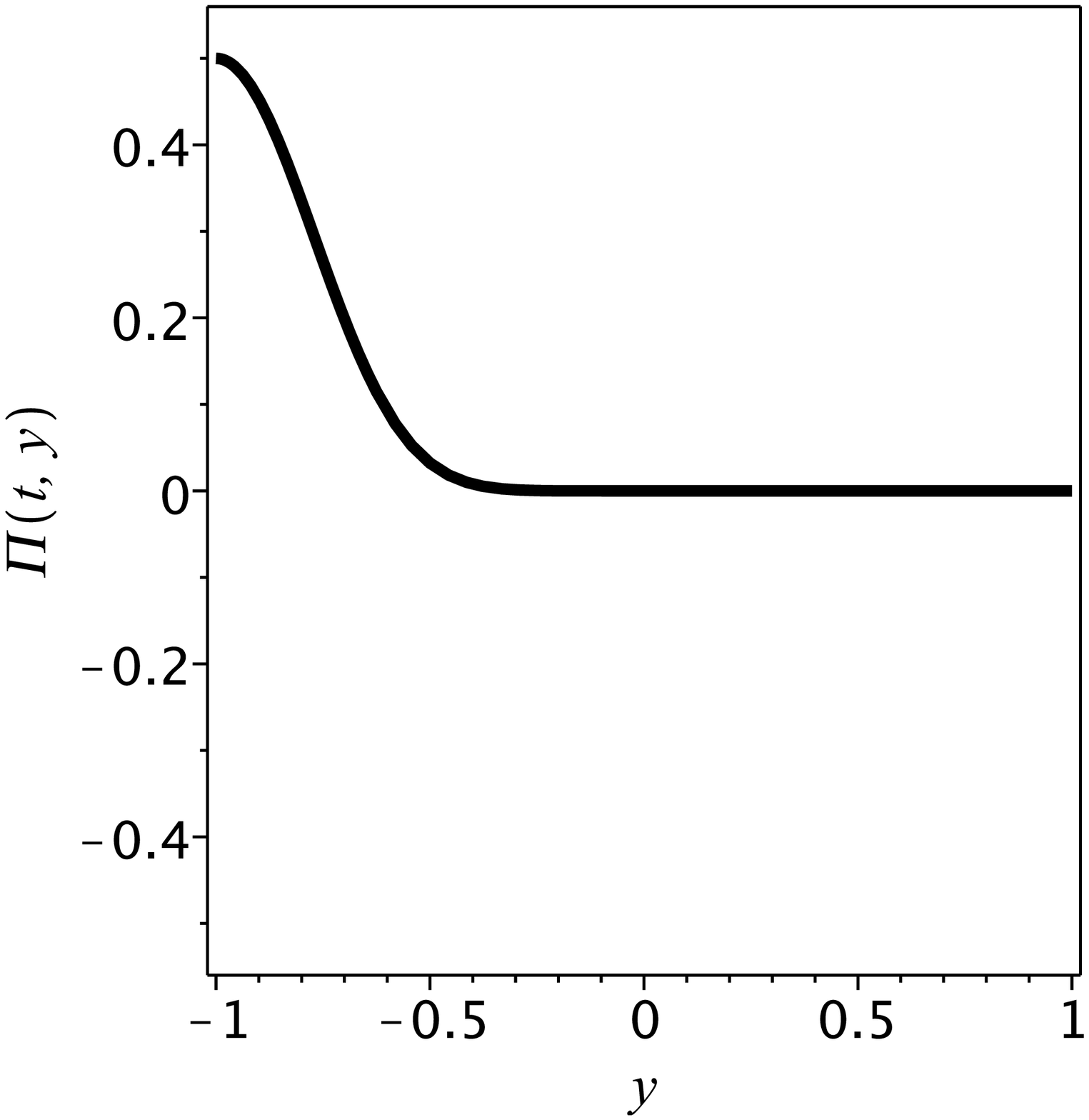}}
\rotatebox{0}{\includegraphics*[scale=0.2]{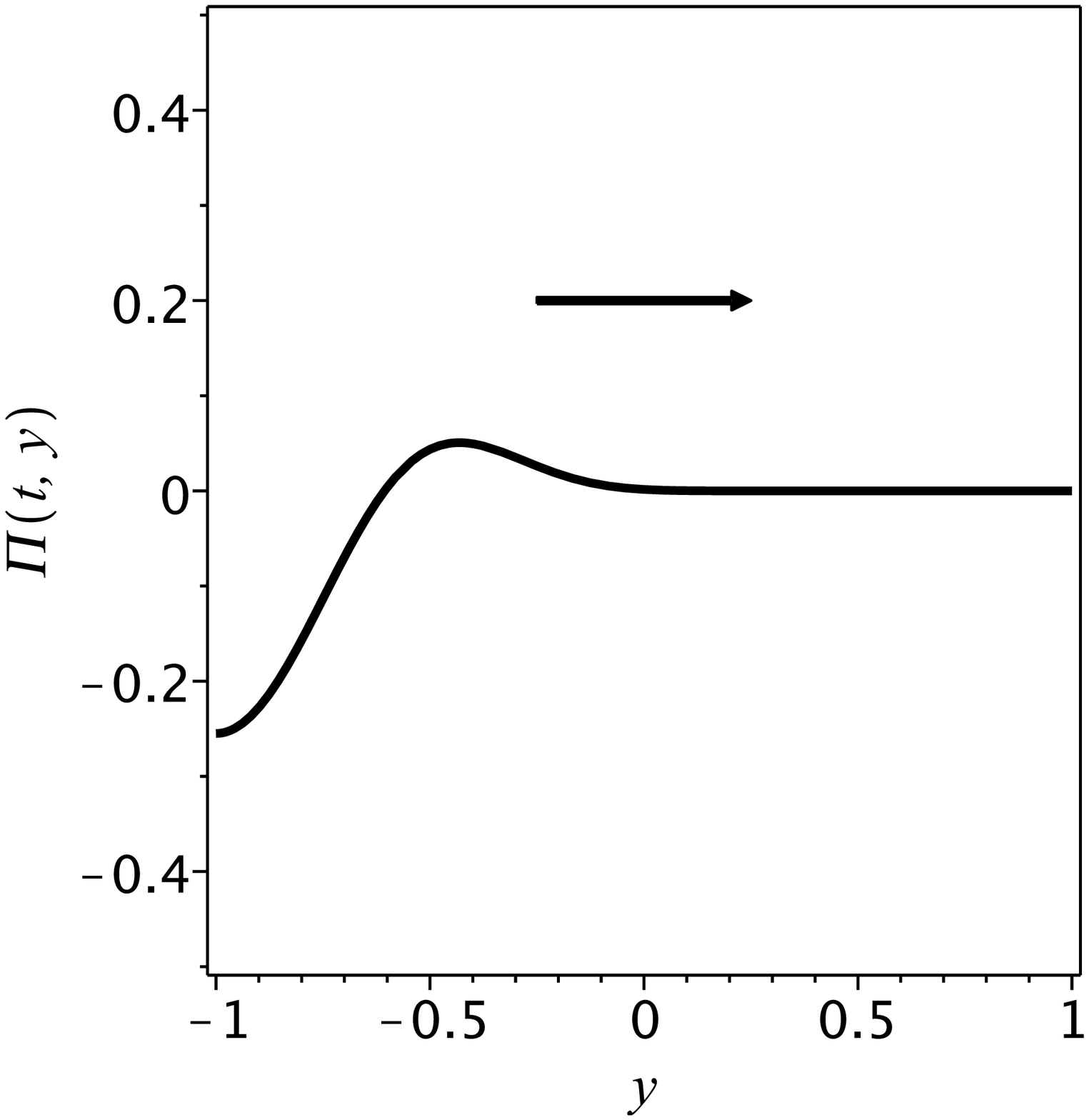}}
\rotatebox{0}{\includegraphics*[scale=0.2]{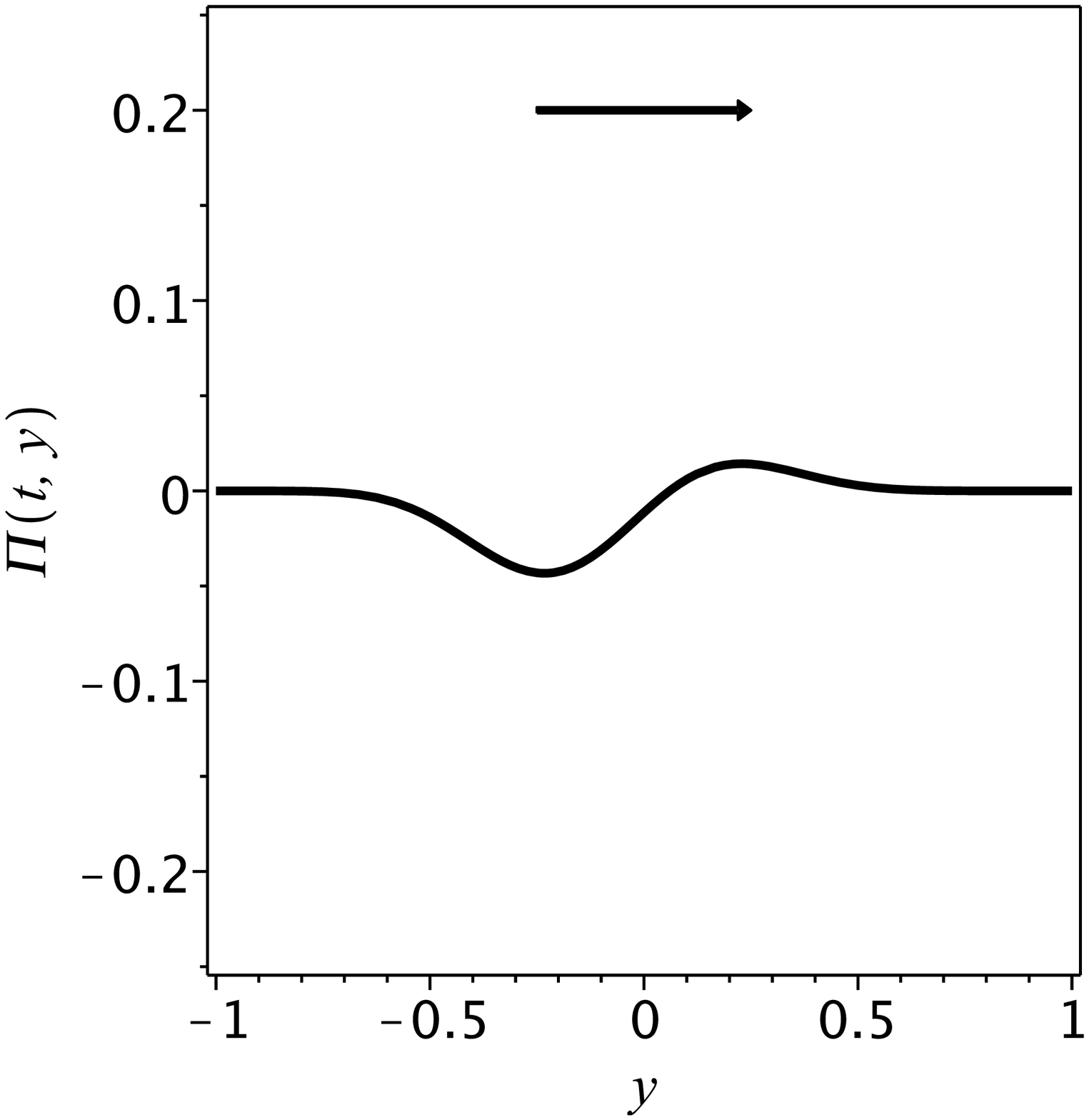}}
\rotatebox{0}{\includegraphics*[scale=0.2]{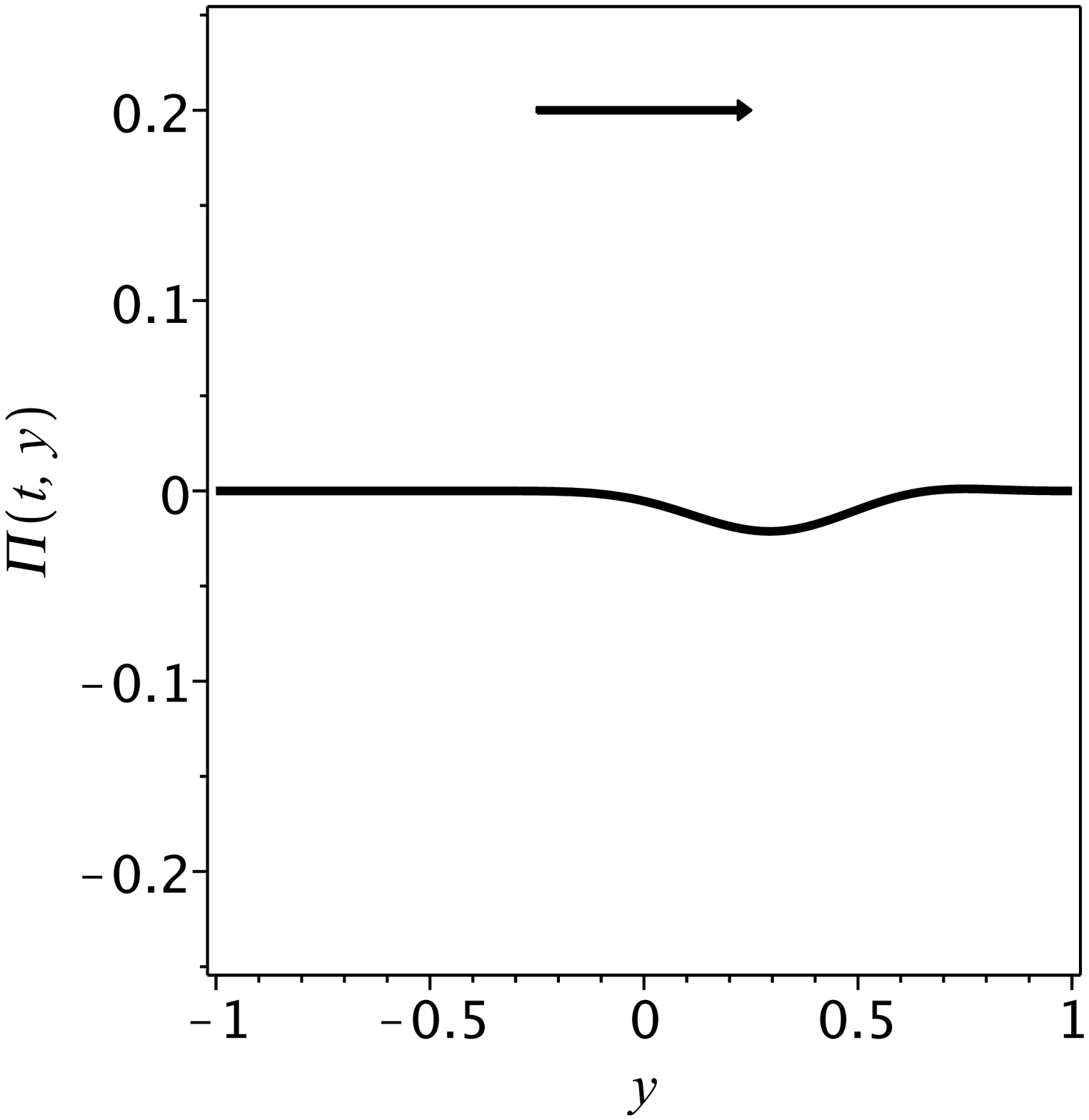}}
\rotatebox{0}{\includegraphics*[scale=0.2]{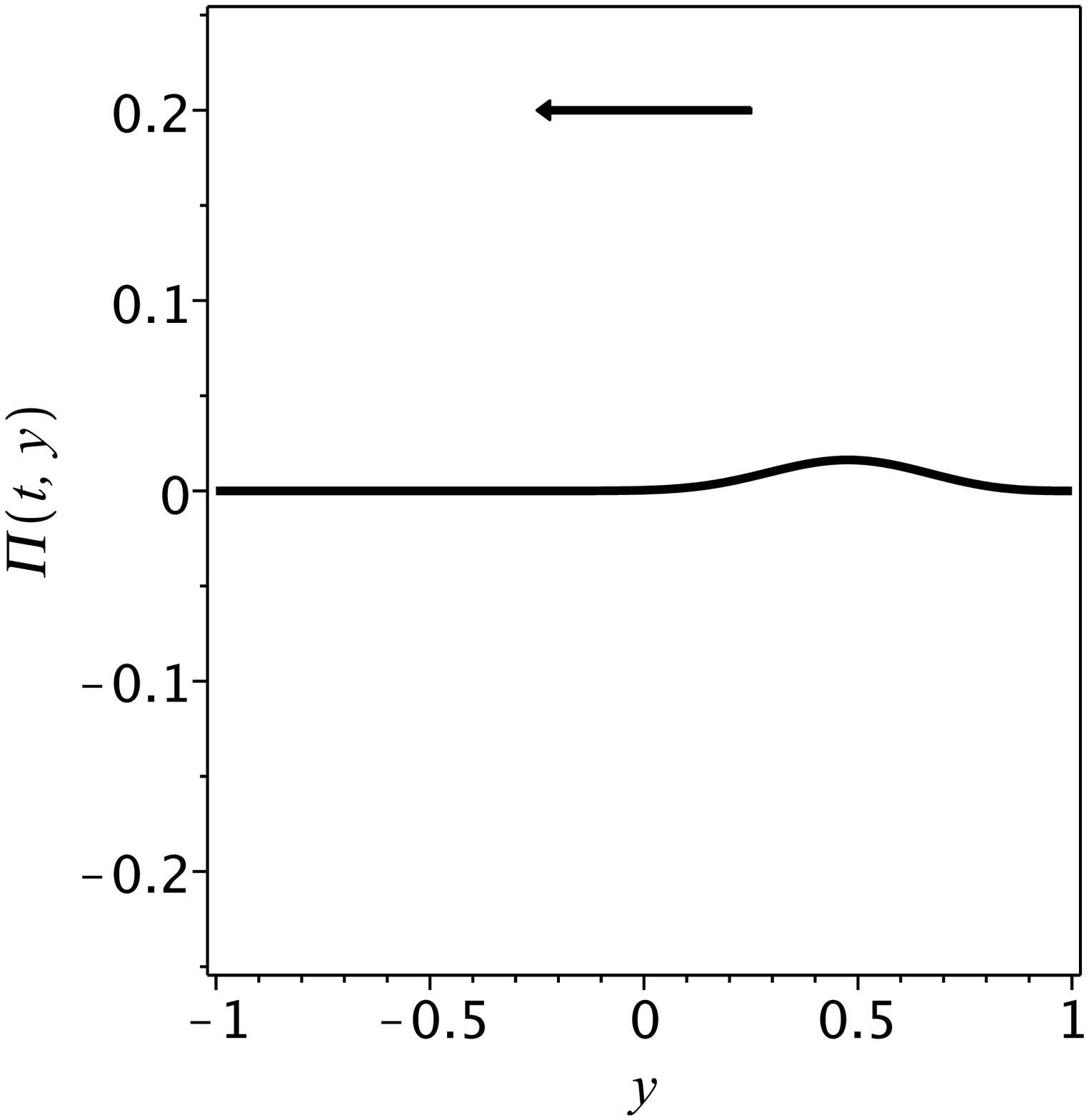}}
\rotatebox{0}{\includegraphics*[scale=0.2]{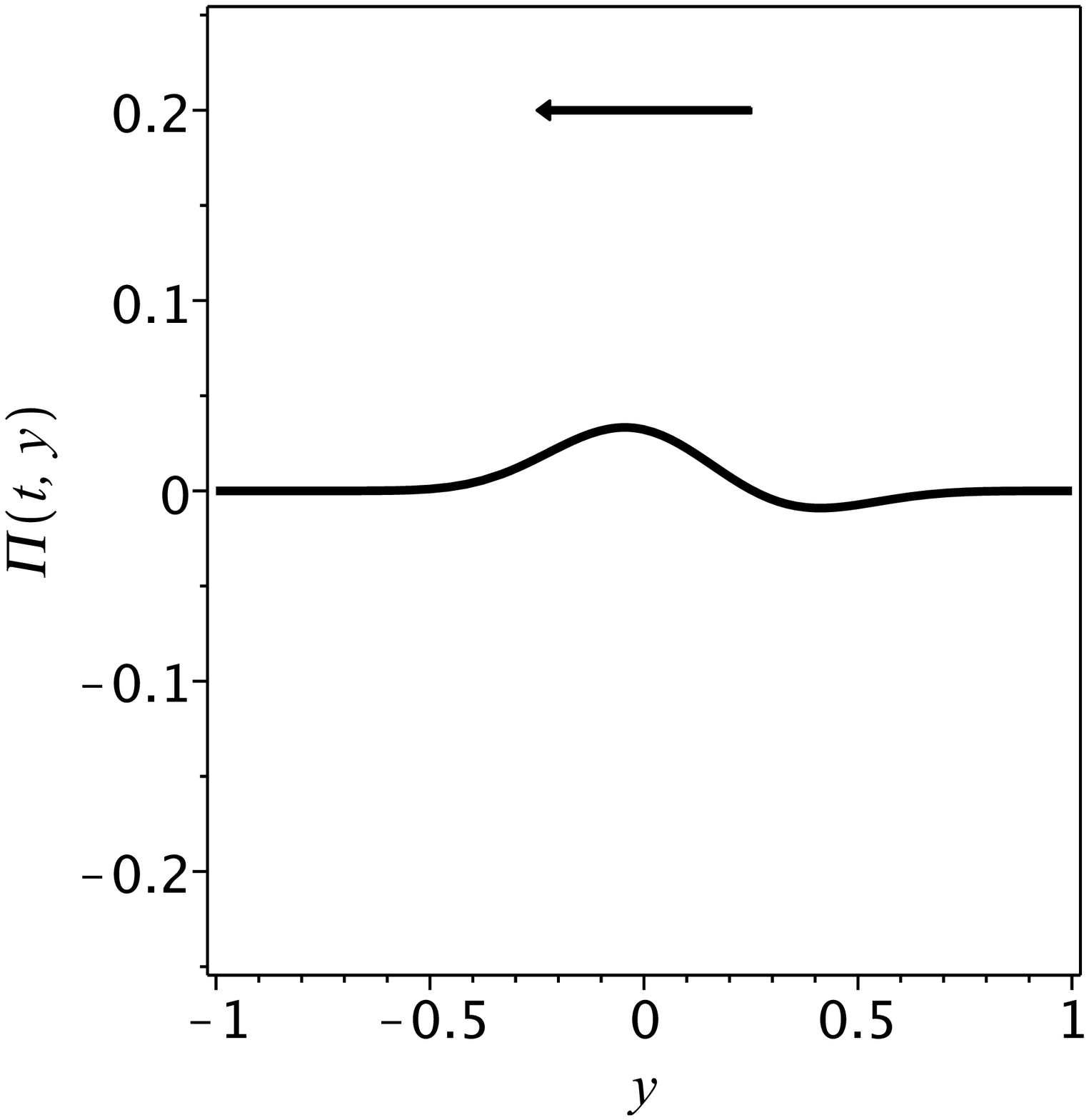}}
\rotatebox{0}{\includegraphics*[scale=0.2]{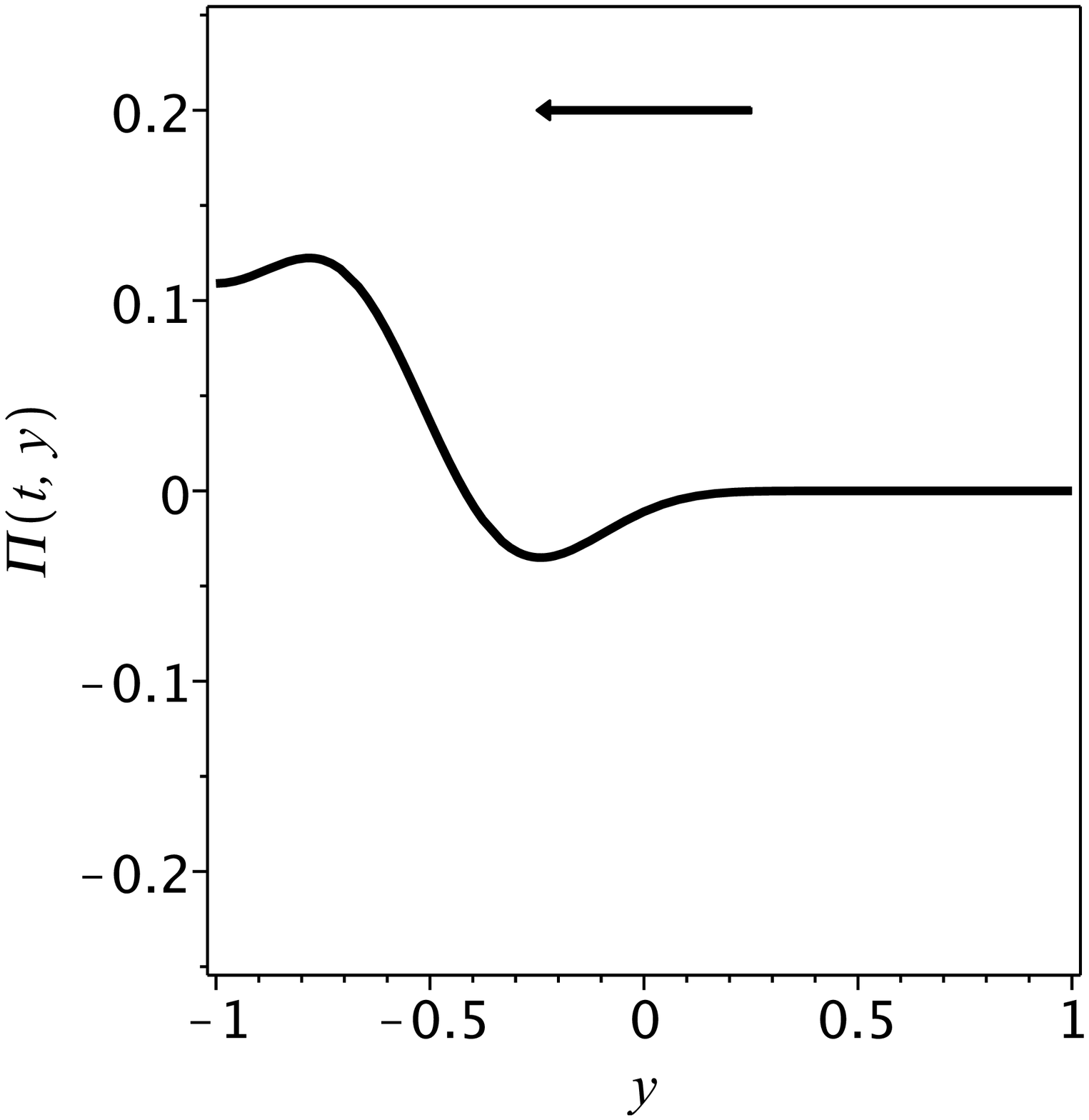}}
\caption{Sequence of snapshots showing the reflection of the scalar field starting from the initial data (first graph). The arrows indicate the sense of propagation and this sequence is about to repeat for a long time. Here the boundaries $y=-1,1$ represent the origin and the spatial infinity. Here $\epsilon_0=0.5$.}
\end{figure}

Let us discuss a convenient way of viewing the dynamics of scalar field in AdS spacetimes. For the sake of simplicity in the weak field regime, $\delta,1-A \ll 1$, the Einstein-Klein-Gordon equation in spectral space can be written as,

\be
\ddot{a}_k(t) - \frac{1}{\tan^2(x(y))}\,[\tan^2(x(y))]^\prime a_k(t) + \mathcal{F}(\delta,1-A,a_k,\dot{a}_k,\ddot{a}_k,..)=0.
\ee


\noindent In this way, the first two terms represent the evolution of the linearized modes (see Wald \cite{wald}) of the scalar field that oscillates in well defined frequencies $\omega_j^2=(3+2j)^2$, $j=0,1,2,...$, and the remaining denoted by $\mathcal{F}$ contains the interaction with the gravitational sector. Therefore, we imagine our system as a set of nonlinear interacting oscillators, and consequently the periodic motion can be understood as taken place on a torus in the abstract phase space spanned by the modes $a_k(t)$. Further numerical experiments indicated that if the term $\mathcal{F}$ is initially small, its influence on the long term dynamics is to alter the natural frequencies of oscillation producing a quasi-periodic motion. Then, the motion becomes unstable and the torus is eventually destroyed forming a strange attractor, which is a signature of chaos in our dynamical system. This means that the evolution of the scalar field in AdS is turbulent. This route to turbulence resembles the Ruelle-Takens scenario \cite{ruelle} where the initial amplitude $\epsilon_0$ plays the role of the control parameter. In Fig. 2, a visualization of this route by a sequence of plots of $\Pi$ and $\varphi$ evaluated at the origin for three values of the initial amplitude, $\epsilon_0=0.5,3.0,5.0$. These curves can be understood as the projections of a trajectory in phase space into the plane $(\Pi,\varphi)$. In the first plot the closed curve means that the motion is periodic as expected since the initial scalar field is very weak ($\epsilon_0=0.5$). By increasing the amplitude the curve is no longer a closed one as a consequence of the new frequencies producing a quasiperiodic motion, and finally for $\epsilon_0=5.0$ the structure is typical of a chaotic system in which the trajectory seems appear to skip around randomly.

The last numerical experiment consists in determining the power spectrum of the scalar of curvature $R=1/l^2[A\cos(x)^2 (-\Pi^2+\Phi^2)-12]
$ evaluated at the origin. We basically have set $\epsilon_0=0.001$ in order to reproduce for a long time the linear behavior, and $\epsilon_0=5.0$. In the region of frequencies under consideration, it is clear the noisy structure of frequencies. Also, a broad band of high frequencies are also present typical of the turbulent regime.


\begin{figure}
\begin{center}
\resizebox{2.1in}{!}{\includegraphics{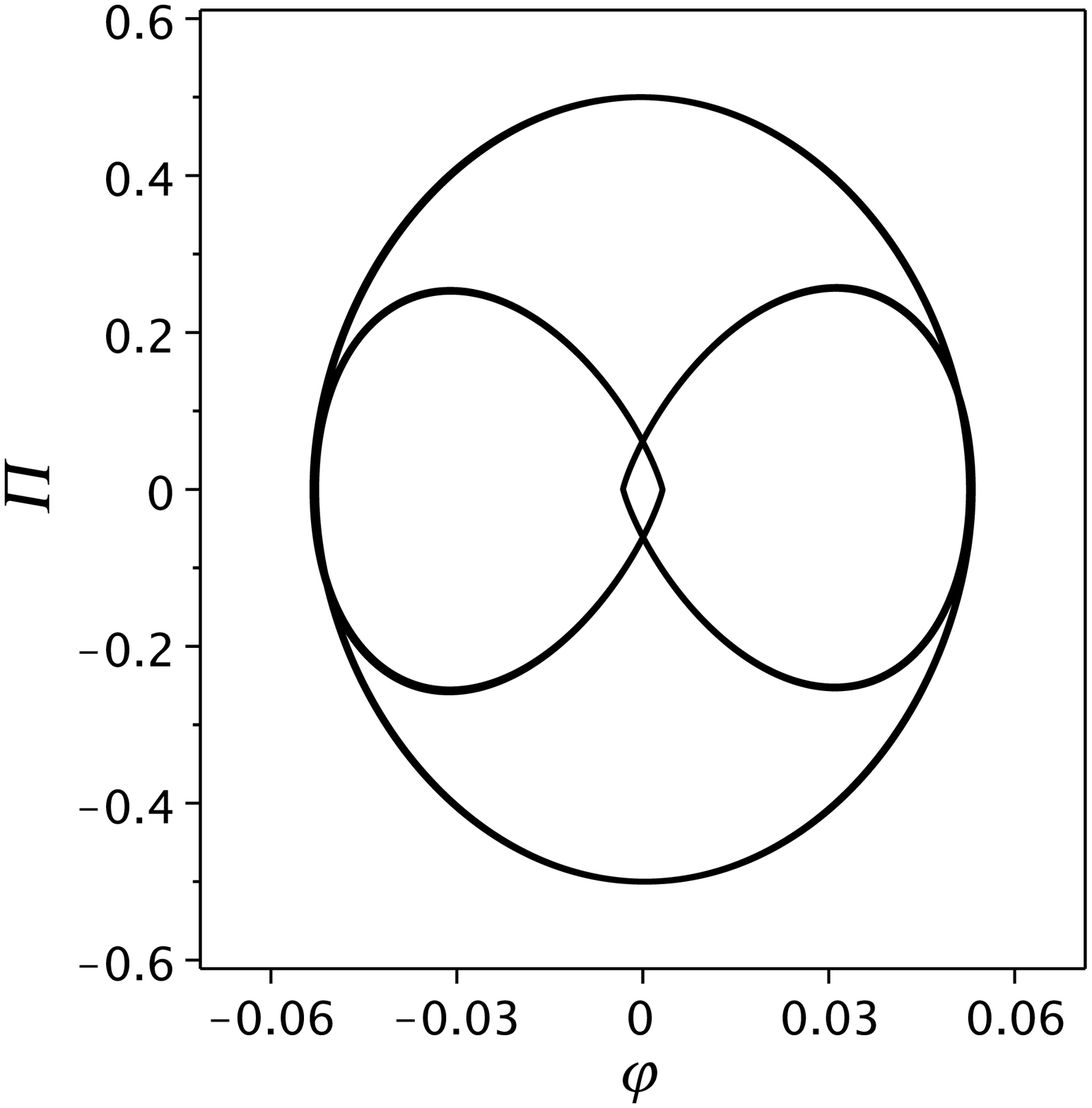}}
\resizebox{2.1in}{!}{\includegraphics{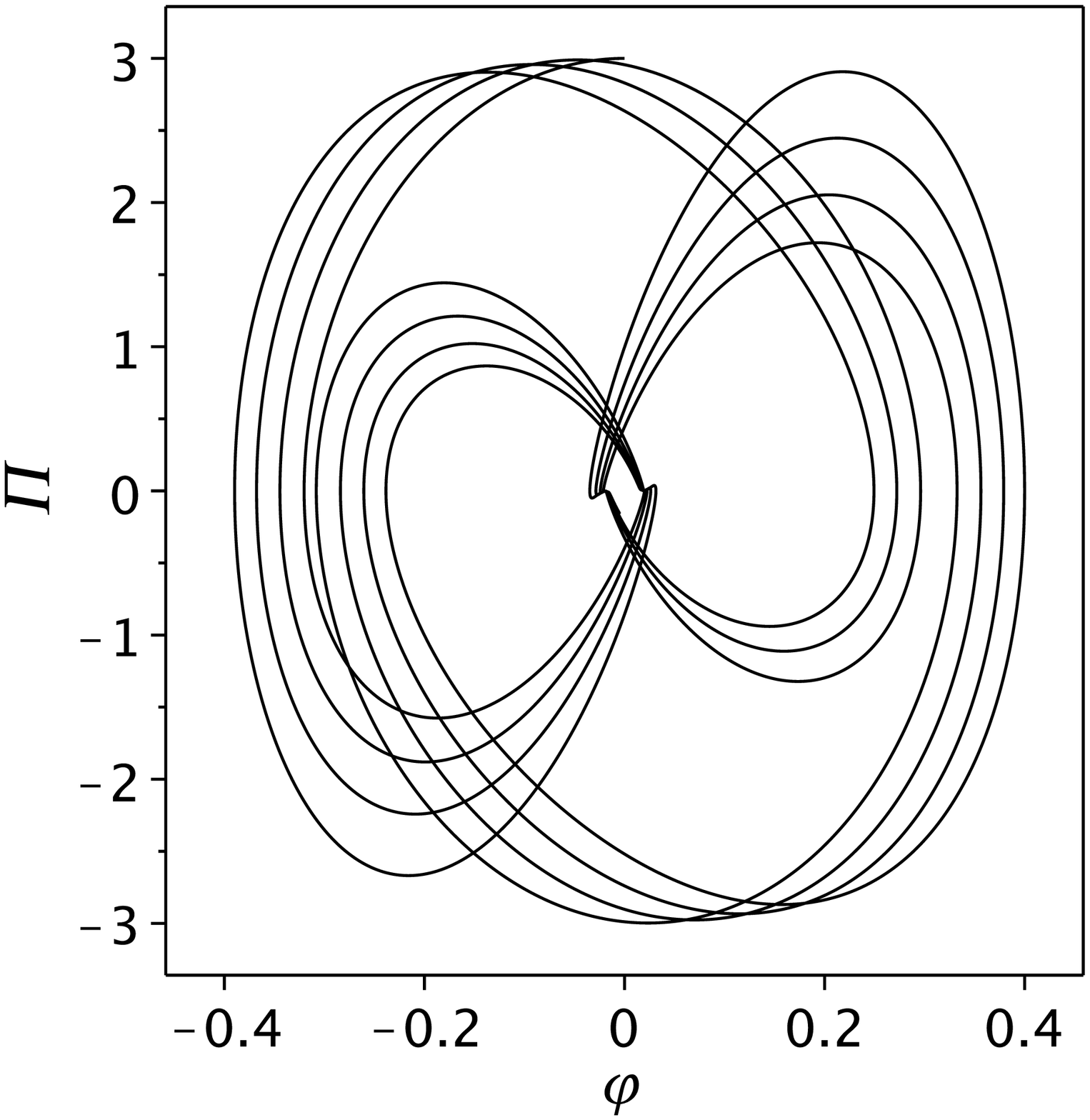}}
\resizebox{2.1in}{!}{\includegraphics{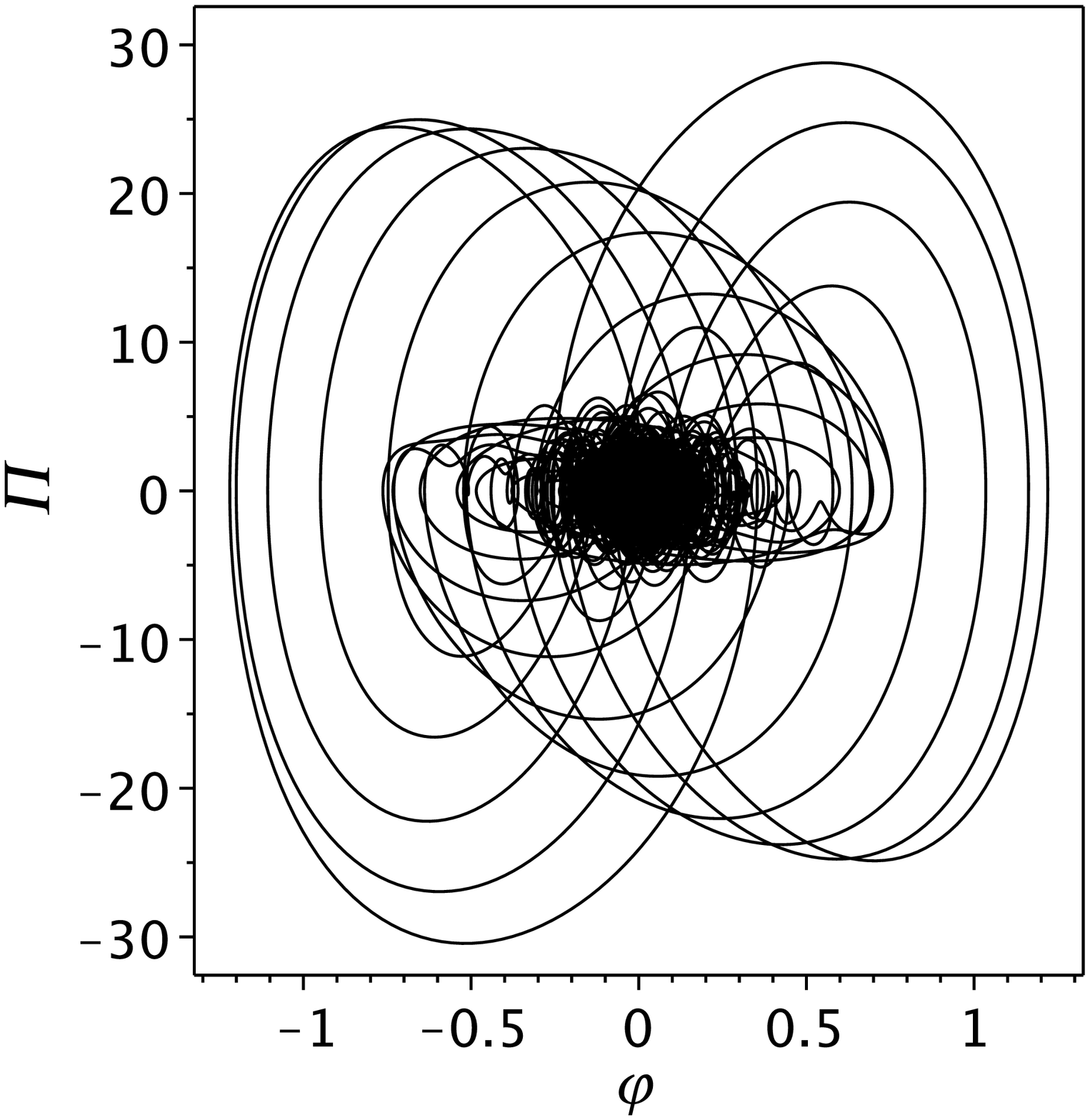}}
\caption{The increase of the initial amplitude $\epsilon_0$ produces the transition from the regular motion on a torus ($\epsilon_0=0.5$) represented by a closed curve on the plane ($\Pi,\phi$) to quasiperiodic motion ($\epsilon_=3.0$), and to turbulent motion ($\epsilon_0=5.0$). In all cases $t_f=30.0$.}
\end{center}
\end{figure}

\begin{figure}
\begin{center}
\rotatebox{0}{\includegraphics*[scale=0.28]{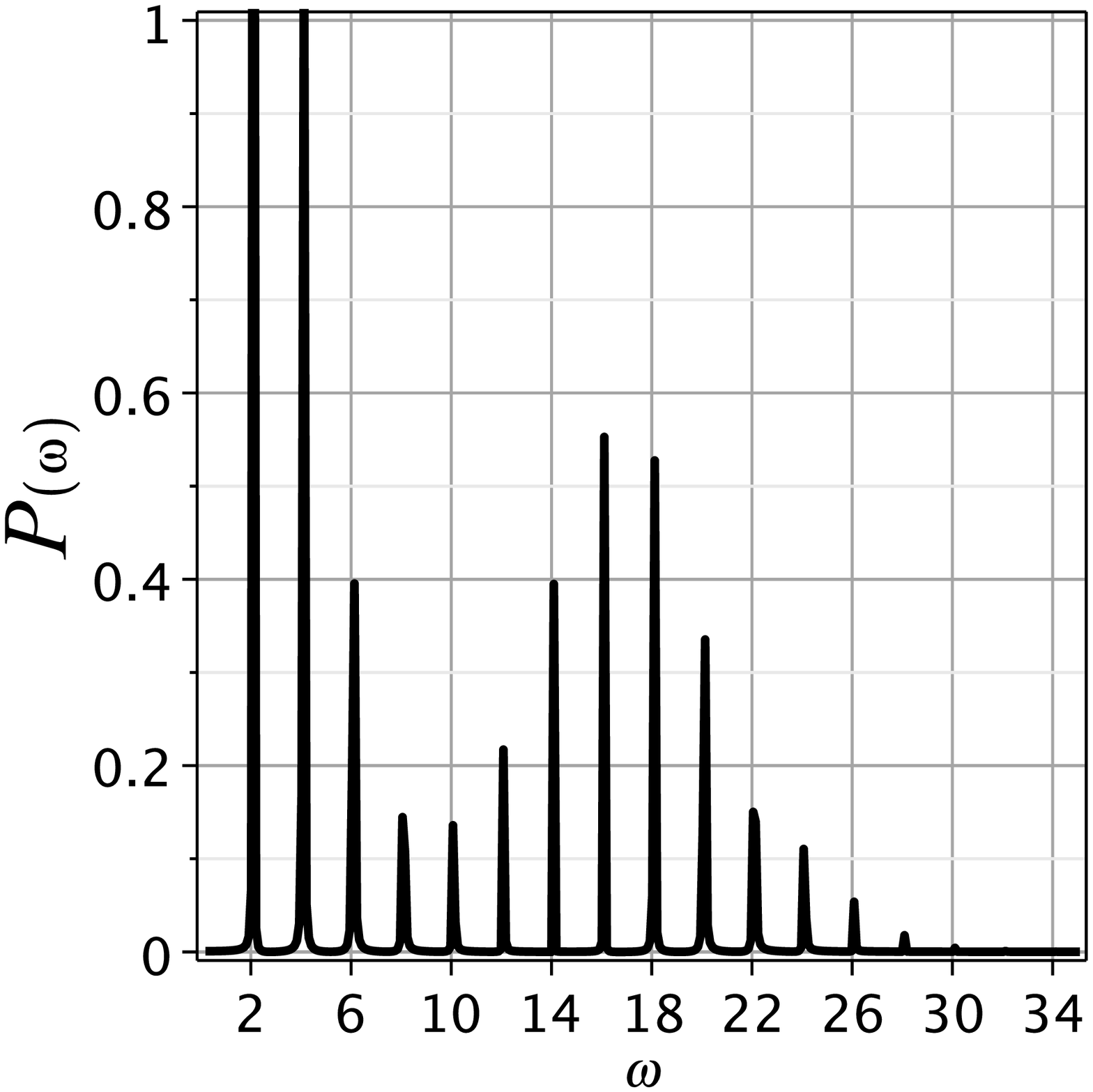}}
\rotatebox{0}{\includegraphics*[scale=0.28]{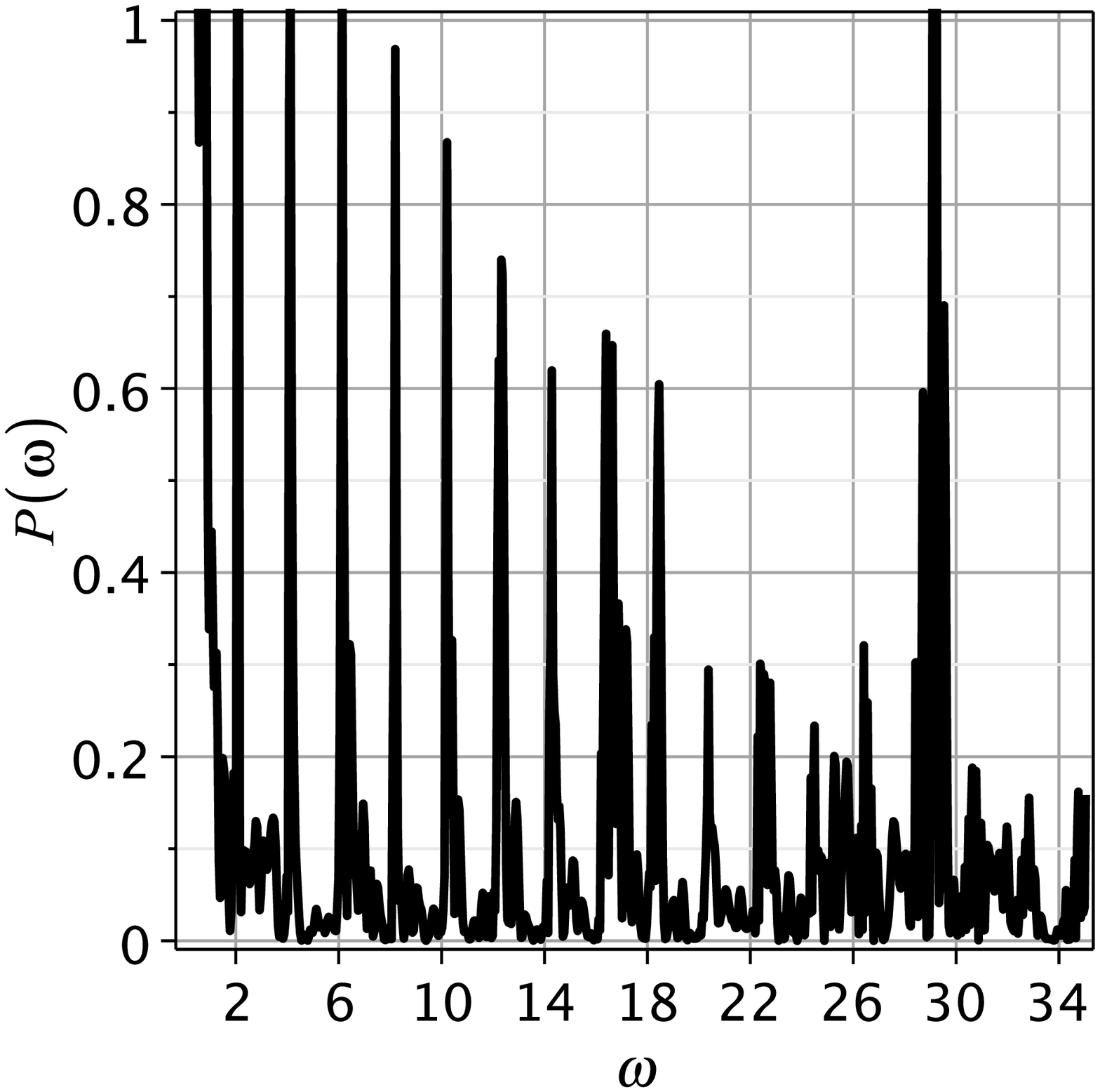}}
\caption{Power spectrum of the scalar of curvature evaluated at the origin for two cases: for very small amplitude $\epsilon_0=0.001$, and for $\epsilon_0=5.0$. In the first case we have the regular or non-chaotic motion with frequencies expected from analytical considerations. In the second case the power spectral is typical of a turbulent system. The broadband frequency is present but not shown here.}
\end{center}
\end{figure}

Many important questions remained unanswered and this only witnesses the fact that the dynamics of of Einstein's equations in asymptotically AdS spacetimes are substantially different from our intuition built on asymptotically Minkowski spacetime.

The result that for an arbitrarily small perturbation the spacetime collapses after a few bounces challenges our intuition and leads to new venues of research into the understanding of the precise mechanism for collapse.

Let us briefly discuss the some of the implications of this result for the dual large $N$ strongly coupled field theories.
There are many arguments leading to the fact that the gravity calculations presented here
pertain to the microcanonical ensemble on the field theory side. First, conservation of energy tells us that we are in a system with
a fixed amount of total energy. Further, since black holes in flat space have negative specific
heat they can cool themselves by transferring energy to the surrounding and consequently be
able to transfer more energy. We expect that small black holes in $AdS$ will behave similarly.
From the field theory point of view sometimes the canonical ensemble is relevant in which
case it is worth revisiting the role of the Hawking-Page phase transition in the context of the
AdS/CFT correspondence. The interpretation of the Hawking-Page transition in the context
of the gauge gravity correspondence leads to the result that below a certain temperature, that
is, below a certain radius of the black hole, the background that is relevant in the thermodynamic
competition is no longer the Schwarzschild black hole but rather thermal $AdS$ \cite{hp}. This
result means that for low temperatures the field theory goes into a confined phase rather than
remaining in the quark gluon plasma. These suggests that the very low temperature black holes
are not relevant from the field theory point of view as they correspond to an unstable phase when viewed in the canonical ensemble. This situations leads to the all important but conceptually difficult question of whether numerical relativity can accommodate a scenario that would correspond, on the field theory side, to the canonical ensemble.

In this essay we have strived to show, by considering a particular example, that the study of gravity in asymptotically $AdS$ spacetimes has many surprises and challenges in store for various communities. We have also tried to argue that the demand is also on methods.  In particular, spectral methods seem more suited to answer questions about evolution of certain modes while finite difference methods will be more efficient in problems where a singularity forms. We have explicitly displayed techniques to study nonlinear dynamical systems with the end of understanding precisely the mechanism by which perturbations in AdS grow to  the point where black hole formation occurs.

H.P.O thanks the financial support of Brazilian agencies CNPq and FAPERJ. L.A.P.Z. thanks D. Garfinkle and D. Reichmann for collaboration on similar matters and to P. Bizon for various patient explanations. This work is  partially supported by Department of Energy under grant DE-FG02-95ER40899 to the University of Michigan.


\begin{thebibliography}{99}



\bibitem{Maldacena:1997re} J.~M.~Maldacena, Adv.\ Theor.\ Math.\ Phys.\  {\bf 2} (1998) 231 [Int.\ J.\ Theor.\ Phys.\  {\bf 38} (1999) 1113]. E.~Witten, Adv.\ Theor.\ Math.\ Phys.\  {\bf 2} (1998) 253.

\bibitem{bizon} P.~Bizon and A.~Rostworowski, ``On weakly turbulent instability of anti-de Sitter space,'' Phys.\ Rev.\ Lett.\  {\bf 107} (2011) 031102 [arXiv:1104.3702 [gr-qc]]. J.~Jalmuzna, A.~Rostworowski and P.~Bizon, ``A comment on AdS collapse of a scalar field in higher dimensions,'' Phys.\ Rev.\  D {\bf 84} (2011) 085021 [arXiv:1108.4539 [gr-qc]].

\bibitem{leo-garf}
  D.~Garfinkle, L.~A.~Pando Zayas and D.~Reichmann,
 ``On Field Theory Thermalization from Gravitational Collapse,''  JHEP {\bf 1202} (2012) 119  [arXiv:1110.5823 [hep-th]].   D.~Garfinkle and L.~A.~Pando Zayas, ``Rapid Thermalization in Field Theory from Gravitational Collapse,''  Phys.\ Rev.\ D {\bf 84} (2011) 066006  [arXiv:1106.2339 [hep-th]].


\bibitem{dias} O.~J.~C.~Dias, G.~T.~Horowitz and J.~E.~Santos,
  ``Gravitational Turbulent Instability of Anti-de Sitter Space,''
  arXiv:1109.1825 [hep-th].

\bibitem{boyd} J. Boyd, \textit{Chebyshev and Fourier spectral methods}, 2nd edn. (Dover, New York, 2001).

\bibitem{wald} A. Ishibashi and R. M. wald, Class. Quant. Grav. \textbf{21}, 2981 (2004).

\bibitem{ruelle} S. Newhouse, D. Ruelle, and F. Takens, Comm. Math. Phys., \textbf{64}, 35 (1978).



\bibitem{hp}
S.~W.~Hawking and D.~N.~Page, ``Thermodynamics of Black Holes in anti-De Sitter Space,''  Commun.\ Math.\ Phys.\  {\bf 87} (1983) 577.   E.~Witten,
  ``Anti-de Sitter space, thermal phase transition, and confinement in gauge theories,''  Adv.\ Theor.\ Math.\ Phys.\  {\bf 2} (1998) 505  [hep-th/9803131].








\end{thebibliography}
\end{document}